\documentclass[12pt]{iopart}

\usepackage{graphics}
\usepackage{epsfig}
\usepackage[colorlinks=false,hypertexnames=false]{hyperref}
\usepackage{hypcap}
\usepackage{subfig}
\usepackage{multirow}
\usepackage{placeins}
\usepackage[usenames,dvipsnames]{color}
\usepackage[table]{xcolor}
\usepackage{algorithm}
\usepackage{algorithmic}
\usepackage[normalem]{ulem}
\usepackage{bbm}
\usepackage{comment}
\hypersetup{pdfborder={0 0 0},pdftitle={}}
\makeatletter
\renewenvironment{figure}[1][\fps@figure]{
\edef\@tempa{\noexpand\@float{figure}[#1]}
\@tempa\capstart
}{
\end@float
}
\makeatother
                                \usepackage[comma, authoryear]{natbib}
\hypersetup{pdftitle={[PET motion correction using real-time MR imaging}}
\usepackage{bm}
\usepackage{iopams}

\renewcommand{\r}{\bm{r}}
\newcommand{\Us}{U_s}
\newcommand{\hUs}{\hat{U}_s}
\newcommand{\Vt}{V_t}

\newcommand{\Fs}{F_s}

\renewcommand{\d}{d}
\newcommand{\z}{z}
\newcommand{\x}{\bm{x}}
\newcommand{\y}{\bm{y}}
\newcommand{\C}{\bm{C}}
\newcommand{\s}{\bm{s}}
\newcommand{\A}{\bm{A}}
\newcommand{\N}{\bm{S}}
\newcommand{\Ns}{\overline{\bm{S}}}
\newcommand{\F}{\bm{F}}
\newcommand{\G}{\bm{G}}
\newcommand{\Gs}{\overline{\bm{G}}}
\newcommand{\M}{\bm{M}}
\date{}

\begin{document}

\title[PET motion correction using real-time MR imaging]{Motion correction for PET using subspace-based real-time MR imaging in simultaneous PET/MR}

\author{%
  Thibault Marin\textsuperscript{* 1,2}, %
  Yanis Djebra\textsuperscript{* 1,2,3}, %
  Paul K. Han\textsuperscript{1,2}, %
  Yanis Chemli\textsuperscript{1,2,3}, %
  Isabelle Bloch\textsuperscript{3}, %
  Georges El Fakhri\textsuperscript{1,2}, %
  Jinsong Ouyang\textsuperscript{1,2}, %
  Yoann Petibon\textsuperscript{1,2}, %
  Chao Ma\textsuperscript{$\dagger$ 1,2}}

\address{\textsuperscript{*} Equal contribution}
\address{\textsuperscript{$\dagger$} Corresponding author}
\address{\textsuperscript{1} Gordon Center for Medical Imaging, Department of Radiology, Massachusetts General Hospital, Boston MA, 02114, USA}
\address{\textsuperscript{2} Harvard Medical School, Boston MA, 02115, USA}
\address{\textsuperscript{3} LTCI, T{\'e}l{\'e}com Paris, Institut Polytechnique de Paris, France}

\begin{abstract}
  Image quality of PET reconstructions is degraded by subject motion occurring
  during the acquisition. MR-based motion correction approaches have been
  studied for PET/MR scanners and have been successful at capturing regular
  motion patterns, when used in conjunction with surrogate signals
  (e.g. navigators) to detect motion. However, handling irregular respiratory
  motion and bulk motion remains challenging. In this work, we propose an
  MR-based motion correction method relying on subspace-based real-time MR
  imaging to estimate motion fields used to correct PET reconstructions. We
  take advantage of the low-rank characteristics of dynamic MR images to
  reconstruct high-resolution MR images at high frame rates from highly
  undersampled k-space data. Reconstructed dynamic MR images are used to
  determine motion phases for PET reconstruction and estimate phase-to-phase
  nonrigid motion fields able to capture complex motion patterns such as
  irregular respiratory and bulk motion. MR-derived binning and motion fields
  are used for PET reconstruction to generate motion-corrected PET images. The
  proposed method was evaluated on in vivo data with irregular motion patterns.
  MR reconstructions accurately captured motion, outperforming state-of-the-art
  dynamic MR reconstruction techniques. Evaluation of PET reconstructions
  demonstrated the benefits of the proposed method over standard methods in
  terms of motion artifact reduction. The proposed method can improve the image
  quality of motion-corrected PET reconstructions in clinical applications.
\end{abstract}

\vspace{2pc}

\submitto{\PMB}

\section{Introduction}
\label{introduction}
Motion, including physiological motion (\emph{i.e.}, cardiac and respiratory motions)
and involuntary bulk motion, is a major source of image quality degradation in
Positron Emission Tomography (PET), which can result in spatial blurring
artifacts and mismatch between emission and attenuation maps, altering
quantification of tracer concentration and deteriorating the diagnostic value of
PET images \citep{Liu2009,Ouyang2013,Rubeaux2017}. The conventional way to
handle motion in PET is the gating method, which bins PET list-mode data to
different cardiac and/or respiratory motion phases followed by reconstructions
of images of each phase. However, gating results in increased noise levels due
to the reduced number of events in each motion phase. To address the
limitations of the gating method, many PET motion correction methods have been
developed, which consist of two consecutive steps: motion field estimation and
motion correction by either applying the estimated motion fields to the gated
images or modeling it within motion-compensated PET image reconstruction
\citep{Rahmim2013}.

PET motion correction methods can be divided into two major subcategories,
depending on how the motion field is estimated: PET-based methods and Magnetic
Resonance (MR)-based methods. In the PET-based motion correction methods, the
measured emission data are first assigned to specific motion phases based on
surrogate signals \citep{Jin2013}, \emph{e.g.}, electrocardiogram (EKG), respiratory
bellow, optical tracking, etc. \citep{Fulton2002,Montgomery2006,Yu2016a}, or the
PET-data themselves \citep{Kesner2009,Sun2019,Lu2019}, \emph{e.g.}, center of mass,
time-of-flight information, frame-by-frame images, etc. Motion fields are then
estimated by registering the reconstructed image of each phase to a reference
phase \citep{Dawood2008}. However, the accuracy of the motion fields estimated
by the PET-based methods is limited by low signal-to-noise ratio (SNR),
especially in the case of dual gating, and the overall lack of anatomical
structural information of PET images \citep{Ouyang2013,Petibon2019}.

The increasing availability of hybrid PET/MR systems provides a unique
opportunity for mitigating effects of motion in PET using MR-based motion
correction. Because of its excellent soft-tissue contrast, high spatial
resolution, and high SNR, MR provides more accurate estimation of motion fields
than the PET-based methods. MR-based PET motion correction methods have been
successfully applied to compensate respiratory and cardiac motion in various
applications involving both static and dynamic PET imaging
\citep{Petibon2013,Huang2014,Petibon2019,Catana2015,Gillman2017,Kustner2017}.
One major limitation of the MR-based motion correction methods is that the
conventional noniterative MR imaging methods are unable to resolve cardiac or
respiratory motion in real time due to the slow imaging speed. Binning-based MR
imaging methods \citep{Grimm2015,Rank2016,Feng2016,Munoz2018,Robson2018} are
often used to address this issue, where MR k-space data are grouped into
different motion phases based on surrogate signals (e.g., EKG), navigator
signals, or k-space data alone, and images of each motion phase are then
reconstructed for the estimation of motion fields. However, the binning-based
MR imaging methods suffer from three noticeable limitations. First, they assume
pseudo periodic motion, which does not hold well in the case of arrhythmia and
irregular respiratory motion. Second, they rely on either surrogate signals or
navigator signals acquired along a single direction to assign k-space data to
specific motion phases, which cannot reliably capture involuntary bulk motion.
Third, their performance is limited by the inherent trade-off between the number
of motion phases (and thus the accuracy of motion field measurement) and data
acquisition time.

In this work, we propose a real-time MR imaging method for PET motion
corrections in PET/MR. High resolution real-time MR imaging is achieved by a
subspace-based imaging method, which takes advantage of a unique property of
high-dimensional dynamic MR signals known as partial separability (PS)
\citep{Liang2007}. The PS-model takes advantages of the spatial-temporal
correlations of dynamic MR images, significantly reduces the number of unknowns
of the underlying spatiotemporal signal, and makes it possible to recover high
resolution, high frame-rate dynamic MR images from highly undersampled k-space
data \citep{Zhao2012,Christodoulou2014}. For PET motion correction, the
reconstructed real-time MR images are used to determine motion phases and
estimate motion fields. PET list-mode data are binned into sinograms
accordingly and ordered-subset expectation-maximization (OSEM) reconstruction
\citep{Hudson1994} is performed integrating the estimated displacement in the
system matrix for motion correction. We demonstrate the performance of the
proposed method by carrying out in vivo \textsuperscript{18}F-FDG PET/MR imaging
experiments using a 3T simultaneous PET/MR scanner.

\section{Methods}
\label{methods}
\subsection{PET/MR imaging experiment}
\label{methods_pet_mr_acquisition}
An \textsuperscript{18}F-FDG PET/MR scan was performed on one healthy subject
under a study protocol approved by our local IRB. PET and MR data were
simultaneously acquired 30 minutes after \textsuperscript{18}F-FDG injection
(around 10 mCi) using a 3T PET/MR scanner (Siemens Biograph mMR, Siemens
Healthcare, Erlangen, Germany).

\begin{figure}[htbp]
\centering
\includegraphics[width=0.9\textwidth]{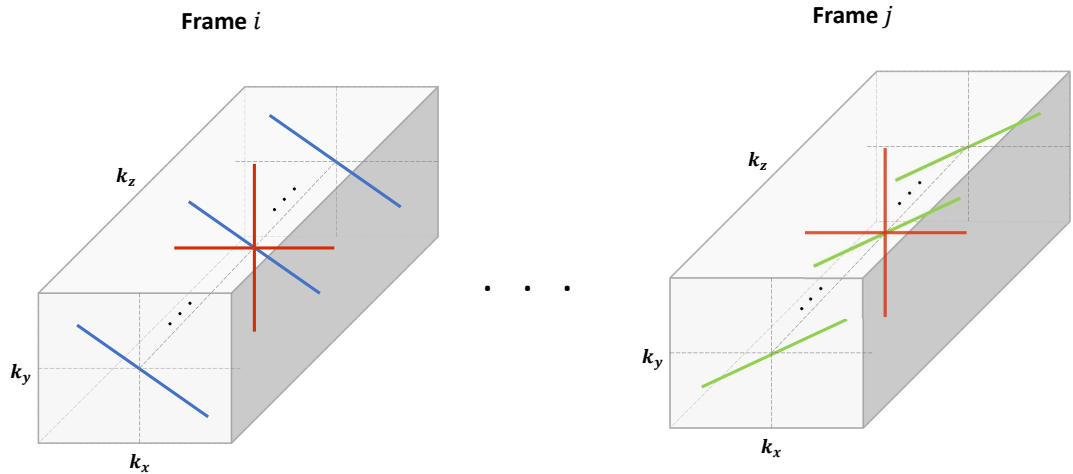}
\caption{\label{fig-mr-sampling}MR Sampling scheme used for the proposed method. 35 lines per frame are acquired: 1 line along \(k_z\) and 34 lines in the \(k_x\)-\(k_y\) plane. For a better visibility, only the \(k_x\)-\(k_y\) in-plane acquired lines are shown in the figure. Two training lines along \(k_x\) and \(k_y\) at \(k_z\) = 0 are consistently acquired through the whole acquisition to estimate the temporal basis \(V_t\) (red lines). A random angle is chosen every frame and is consistently acquired every \(k_z\) for imaging.}
\end{figure}

Two 5-minute MR acquisitions were performed using a spoiled gradient-recalled
echo (GRE) sequence with stack-of-stars radial sampling trajectories in the
coronal plane. The imaging parameters are as follows: image size~\(= 384
\times 384 \times 32\), resolution~\(= 1.9 \times 1.9 \times
5\,\mathrm{mm}^3\), TR/TE~=~3/1.6 ms, and flip angle = 7 degrees.
The (k,t)-space data were acquired using a random sampling pattern shown in
figure~\ref{fig-mr-sampling}. A total of 35 k-space spokes were sampled in
each frame, resulting in a frame rate of 9.5 volumes per second. For each
frame, the first 3 spokes were respectively acquired along the \(k_x\), \(k_y\)
and \(k_z\) direction across the center of the k-space to estimate the temporal
basis of the partially separable (PS) model detailed next. The remaining 32
spokes were along a random angle in the \(k_x\)-\(k_y\) plane for every \(k_z\).
During the first 5-minute acquisition, the subject was instructed to move once
to assess the effect of both respiratory and bulk motion. During the second
5-minute acquisition, the subject was instructed to simulate an irregular
respiratory pattern including both deep and shallow breaths. The
vendor-provided two-point Dixon sequence was performed with breath-holding to
obtain attenuation coefficients.

\subsection{Subspace-based image reconstruction}
\label{methods_low_rank_reconstruction}
Denote the dynamic image series as \(\rho(\r, t)\) and its matrix representation
\(\C \in \mathbbm{C}^{N\times M}\) such that:
\begin{equation}
  \C = \left[\begin{array}{ccc}
              \rho(\r_1, t_1) & \ldots & \rho(\r_1, t_M)\\
              \vdots          & \ddots & \vdots\\
              \rho(\r_N, t_1) & \ldots & \rho(\r_N, t_M)
            \end{array}
          \right].
\end{equation}
We express \(\rho(\r, t)\) as a PS model \citep{Liang2007}:
\begin{equation}
\rho(\r, t) = \sum_{l=1}^{L} u_{l}(\r) v_{l}(t),
\end{equation}
or, equivalently \(C\) as:
\begin{equation}
\C = \Us \, \Vt,
\end{equation}
where \(\Vt \in \mathbbm{C}^{L\times M}\) concentrates in rows the temporal
basis function \(v_{l}\) of the PS model up to order \(L\) and \(\Us \in
\mathbbm{C}^{N\times L}\) concentrates in columns the corresponding spatial
coefficients \(u_{l}\).

We estimate the temporal basis functions using the training data acquired at
every frame. Assuming \(p\) training lines, we form the so-called Casorati
matrix \(\C_t \in \mathbbm{C}^{p N_f \times M}\) by stacking the signal of the
\(p\) training lines at each frame, where \(N_f\) is the number of samples
acquired in each k-space line. The temporal basis functions
\(\{v_{l}\}_{l=1}^{L}\) can then be estimated by calculating the first \(L\) right
eigenvectors of \(\C_t\) using Singular Value Decomposition (SVD).

Once \(\Vt\) is obtained (denoted by \(\hat{\Vt}\)), the image reconstruction
problem is reduced to the determination of the spatial coefficients matrix
\(\Us\). We solve this problem by fitting the PS model to the undersampled
(k,t)-space data with additional sparsity constraints \citep{Zhao2012}:
\begin{equation}
\label{eq-cost-mr}
  \hUs = \arg\min_{\Us}
  \left\|\d - \Omega\left(\Fs \Us \hat{\Vt} \right)\right\|_2^2 +
  \lambda_1 \left\|T(\Us \hat{\Vt})\right\|_1 + \lambda_2 \left\|\Us\right\|_F,
\end{equation}
where \(d\) is the measured k-space data, \(\Fs\) is the Fourier transform
operator in the spatial domain, \emph{i.e.} Non-Uniform FFT (NUFFT) operator
\citep{Fessler2003} for the stack-of-stars trajectory, \(\Omega\) is the sparse
sampling operator in the (k,t)-space, \(T\) is the finite difference operator
along both the spatial and temporal directions, \(\left\|.\right\|_F\) is the
Frobenius norm and the scalar variables \(\lambda_1\) and \(\lambda_2\) are
regularization parameters. The first term in Eq.~(\ref{eq-cost-mr}) is a data
fidelity term, the second term promotes sparsity in the reconstructed image and
the third term favors minimal norm solutions for \(\Us\).

We solve the optimization problem in Eq.~(\ref{eq-cost-mr}) using the
Alternating Direction Methods of Multipliers (ADMM) algorithm \citep{Boyd2011},
which leads to solving the following three sub-optimization problem in an
alternative fashion:
\begin{eqnarray}
  \fl\z^{(k + 1)} = \mathcal{S}_{\frac{\lambda_1}{\mu}}\left(
  T\left(\Us^{(k)}\,\hat{\Vt}\right) + \eta^{(k)}\right),
  \label{eq-admm-z}\\
  \fl\Us^{(k + 1)} = \arg\min_{\Us}
  \frac{1}{2} \left\|d - \Omega \left(
  \Fs \Us \hat{\Vt}\right)\right\|_2^2 +
  \frac{\mu}{2} \left\|T\left(\Us \hat{\Vt}\right) - \z^{(k + 1)} +
  \eta^{(k)}\right\|_F^2\nonumber\\ + \, \lambda_2 \left\|\Us\right\|_F,
  \label{eq-admm-us}\\
  \fl\eta^{(k + 1)} = \eta^{(k)} + \left(T\left(\Us^{(k + 1)}\,\Vt\right) -
  \z^{(k + 1)}\right),
\end{eqnarray}
where \(\z\) is the split variable, \(\eta\) is the dual variable, and \(\mu\)
is a scalar relaxation parameter. The \(\z\) update
(Eq.~(\ref{eq-admm-z})) is a soft thresholding operation and the \(\Us\)
update (Eq.~(\ref{eq-admm-us})) is a convex-optimization problem, which
is solved using the conjugate gradient algorithm.

For comparison, we reconstructed MR images using the same data by a binning-based
method, known as XD-GRASP \citep{Feng2016}. The respiratory motion signal used
for binning was processed the same way \citeauthor{Feng2016} did: the Fourier
transform of the data at the center of the k-space at each frame was sorted into
a 2D matrix, with data from each coil concatenated along the first dimension. A
Principal Component Analysis (PCA) was then applied on this matrix and the
component with the highest peak in the respiratory frequency range (0.1, 0.5 Hz)
was selected as the binning signal. The k-space data were then regrouped into 6
balanced bins, \emph{i.e.}, each bin containing the same number of spokes. The
XD-GRASP reconstruction was performed the same way as in \citep{Feng2016}.

\subsection{Motion estimation}
\label{methods_motion}
The reconstructed real-time MR images were first binned into a small number of
phases corresponding to different body positions (respiratory and bulk motion
phases). Binning was performed in three steps. The first step consists in
visually determining the bulk motion phases from the MR images and discarding
time frames corresponding to the transition between bulk motion phases. In the
second step, a bin is assigned to each (real-time) frame by tracking the tip of
the right lobe of the liver over time while ensuring balanced bins (\emph{i.e.}, all
bins should contain a similar number of frames). Finally, a combined MR image
is formed for each bin by averaging all real-time images in a bin. Volumetric
image registration was then performed between all bins and a reference bin using
the multiscale B-spline registration algorithm described in \citep{Chun2009}.

\subsection{PET reconstruction}
\label{methods_pet_reconstruction}
The acquired list-mode PET events were first rearranged into \(B\) sinograms \(\y =
(\y_1, \ldots, \y_B)\) following the binning determined from MR images and
discarding PET list-mode events occurring during bulk motion transitions.

PET reconstruction was performed using the OSEM algorithm \citep{Hudson1994}
integrating the estimated motion fields in the forward model
\citep{Liu2011,Petibon2016}. Let \(\x\) denote the PET image to reconstruct
arranged in vector form. The system matrix, denoted by \(\F\) is decomposed as
\(\F = \N\,\A\,\G\,\M\), where:
\begin{itemize}
\item \(\M = \left[\M_1, \ldots, \M_B\right]^\top\) is a stack of deformation
operators estimated using the procedure described in
Section~\ref{methods_motion},
\item \(\G = \mathrm{diag}(\Gs, \ldots, \Gs)\) is a block-diagonal geometrical
projection matrix constructed by repeating the static projection matrix \(\Gs\)
implemented using Siddon's algorithm \citep{Siddon1985a},
\item \(\A = \mathrm{diag}(\A_1, \ldots, \A_B)\) is a diagonal matrix with
time-varying attenuation coefficients,
\item \(\N = \mathrm{diag}(\Ns, \ldots, \Ns)\) is a diagonal matrix with detector
sensitivity coefficients \(\Ns\) repeated for all bins.
\end{itemize}

With these notations, the motion-corrected OSEM update for a given subset \(l\) is
given by:
\begin{equation}
  \x^{(n + 1)} = \frac{\x^{(n)}}{\F_l^\top \mathbbm{1}}
  \F_l^\top \frac{\y_s}{\F_l \x^{(n)}  + \s_l},
\end{equation}
where \(\F_l\) is the system matrix for the \(l\)-th subset and \(\s_l\) is the
combined additive correction sinogram for subset \(l\) including randoms and
scatter. Correction sinograms were constructed as follows. Random coincidences
were estimated using the delayed window method. Scatter was estimated using the
single scatter simulation algorithm \citep{Werling2002} from an initial
reconstruction performed without motion correction. Scatter was estimated
separately for each bulk motion phase. Attenuation coefficients were obtained
from a vendor-provided Dixon sequence during breath-holding. The attenuation
map was deformed to each bin and forward projected to calculate sinogram-domain
attenuation coefficients.

The OSEM used 12 subsets and 5 iterations. This motion-corrected reconstruction
is denoted by MC in the rest of the paper. For comparison, two other
reconstruction methods were considered: a traditional OSEM without motion
correction (NMC) and a gated reconstruction where only list-mode events
occurring in a given motion phase are reconstructed without motion correction
(Gated). Both reference methods used 4 iterations to account for the difference
in convergence speed, aiming to match the noise level in MC and NMC
reconstructions.

\subsection{Quantitative analysis}
\label{methods_quantitative_analysis}
\begin{figure}[htbp]
\centering
\includegraphics{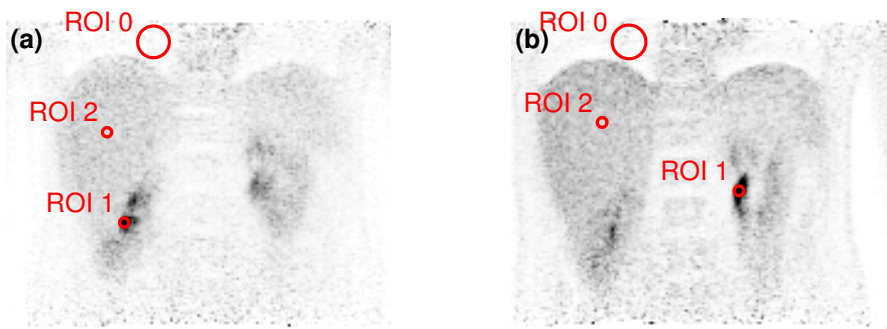}
\caption{\label{fig-pet-exp23-roi}Regions of interest used for quantitative analysis: (a) bulk motion experiment, (b) irregular respiratory motion experiment. ROI 0 in the lung, ROI 1 in the kidney, ROI 2 in the liver.}
\end{figure}

In order to compare PET reconstructions, two evaluation measures were used: the
contrast-to-noise ratio (CNR) and target-to-background ratio (TBR). The
contrast-to-noise ratio is defined as:
\begin{eqnarray}
\label{eq_cnr}
\mathrm{CNR}(x, \mathcal{R}) = \frac{\bar{x}_{\mathcal{R}} -
                                     \bar{x}_{\mathcal{R}_0}}{\sigma_0},
\end{eqnarray}
where \(\sigma_0\) is the standard deviation in the background region
\(\mathcal{R}_0\) (located in the lung) and \(\bar{x}_{\mathcal{R}}\) is the average
activity of image \(x\) in region \(\mathcal{R}\). With the same notations, the
target-to-background ratio is given by:
\begin{eqnarray}
\label{eq_tbr}
\mathrm{TBR}(x, \mathcal{R}) = \frac{\bar{x}_{\mathcal{R}}}
                                    {\bar{x}_{\mathcal{R}_2}},
\end{eqnarray}
where \(\mathcal{R}_2\) is a region located in the liver. Both metrics were
evaluated in a small region located in the kidney (shown in
figure~\ref{fig-pet-exp23-roi}(a)). In the absence of ground truth, the
contrast-to-noise and target-to-background ratios were used as indicators of
image quality.

\section{Results}
\label{results}
\subsection{Correction of bulk motion}
\label{results_experiment_2}
\begin{figure}[htbp]
\centering
\includegraphics[width=0.9\textwidth]{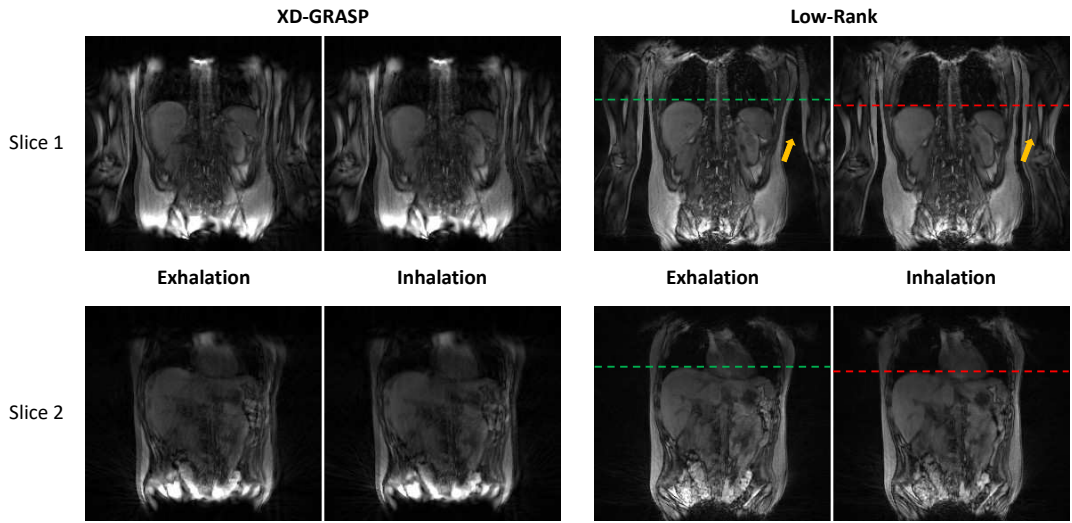}
\caption{\label{fig-mr-exp2-comp}Representative MR images obtained with the XD-GRASP and the proposed method. The red dashed-line indicates the top of the liver position for end-inhalation and the green dashed-line indicates the top of the liver position for end-exhalation. Note that, besides respiratory motion, bulk motion indicated by the yellow arrows is clearly seen in the proposed low-rank based image reconstruction. The images obtained by XD-GRASP show blurring artifacts largely because the bulk motion was not detected from the navigator signal (see figure~\ref{fig-mr-exp2-vt} for more details).}
\end{figure}

\begin{figure}[htbp]
\centering
\includegraphics[width=0.9\textwidth]{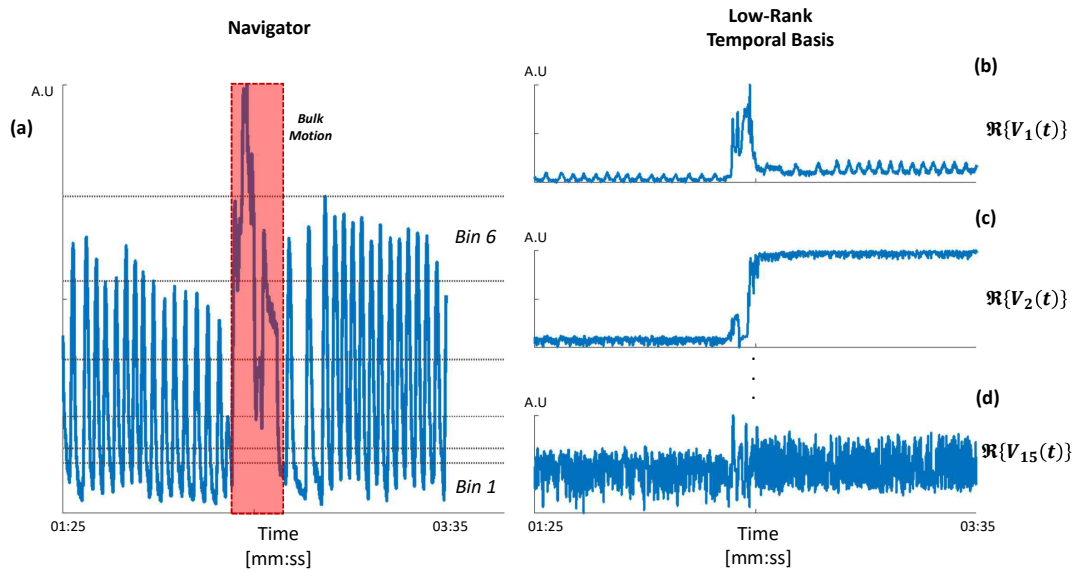}
\caption{\label{fig-mr-exp2-vt}(a) Plot of the processed navigator along time, which was used to bin the k-space data in XD-GRASP. (b) to (d) Real part of the temporal basis \(\Vt\) for the component 1, 2, and 15 of the PS model, respectively.}
\end{figure}

In this experiment the subject was instructed to move after around 2.5 minutes
in the 5-minute acquisition. Images reconstructed by XD-GRASP and the proposed
method at the end inhalation and end-exhalation phases are shown in
figure~\ref{fig-mr-exp2-comp}. The images obtained by XD-GRASP method show
noticeable blurring artifacts largely because the bulk motion was not detected
from the navigator signal. More specifically, figure~\ref{fig-mr-exp2-vt}(a)
shows the navigator signal obtained from the training line along the \(k_z\)
direction in each frame as in \citep{Feng2016}. Since the bulk motion of the
subject was along the x direction (left to right), the navigator signal only
recorded abnormal changes during the bulk motion, \textit{i.e.}, the red region
in figure~\ref{fig-mr-exp2-vt}(a), but did not contain sufficient information
to indicate what type of motion occurred. Therefore, six motion bins were chosen
in XD-GRASP while the k-space data acquired in the red region of
figure~\ref{fig-mr-exp2-vt}(a) were discarded, resulting in blurring artifacts.

\begin{figure}[htbp]
\centering
\includegraphics[width=0.9\textwidth]{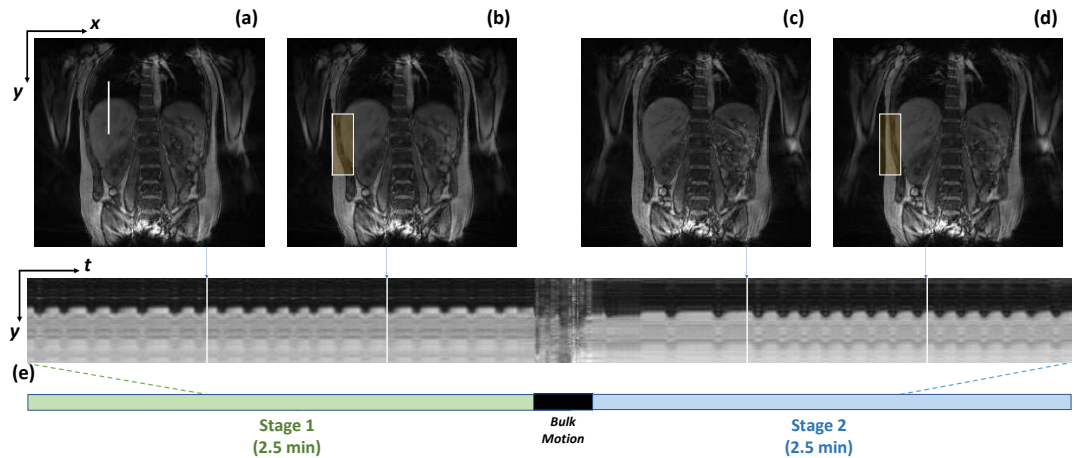}
\caption{\label{fig-mr-exp2-prof}Dynamic MR images reconstructed by the proposed method. The real-time profile (e) is plotted for a part of the experiment where the bulk motion happens, and two representative images (a) \& (b) and (c) \& (d) are shown for each body position. The white line in (a) shows where the time profile was taken. A yellow box at the edge of the patient has been drawn for the first body position (b) and the same box was also drawn for the second body position (d) at the same coordinates (regarding the image). One can clearly see that the body of the subject moved to the right of the image during the bulk motion, and that the proposed method managed to catch that motion.}
\end{figure}

The images obtained by the proposed method shown in
figure~\ref{fig-mr-exp2-comp} successfully capture both respiratory motion (as
indicated by the red and green dashed lines) and bulk motion (as indicated by
the yellow arrows). Figure~\ref{fig-mr-exp2-vt}(b) to (d) show the temporal
basis functions of the PS model estimated from the three training lines, where,
intuitively, the first component (figure~\ref{fig-mr-exp2-vt}(b)) shows
respiratory patterns and the second component (figure~\ref{fig-mr-exp2-vt}(c))
indicates bulk motion. To further demonstrate the real-time capabilities of the
proposed method, figure~\ref{fig-mr-exp2-prof} shows images at multiple time
frames along with a 1D profile through the liver along time. The images from
before and after bulk motion demonstrate the ability to capture both respiratory
and bulk motion. The yellow overlay emphasizes the body displacement between
Stage 1 and Stage 2. The profile plot shows the respiratory motion, captured
for both bulk motion phases. The transition portion between the two bulk motion
phases corresponds to the frames that were discarded in the PET reconstruction.
Videos showing the reconstructed MR images are available in the Supplementary
Material M1.

\begin{figure}[htbp]
\centering
\includegraphics{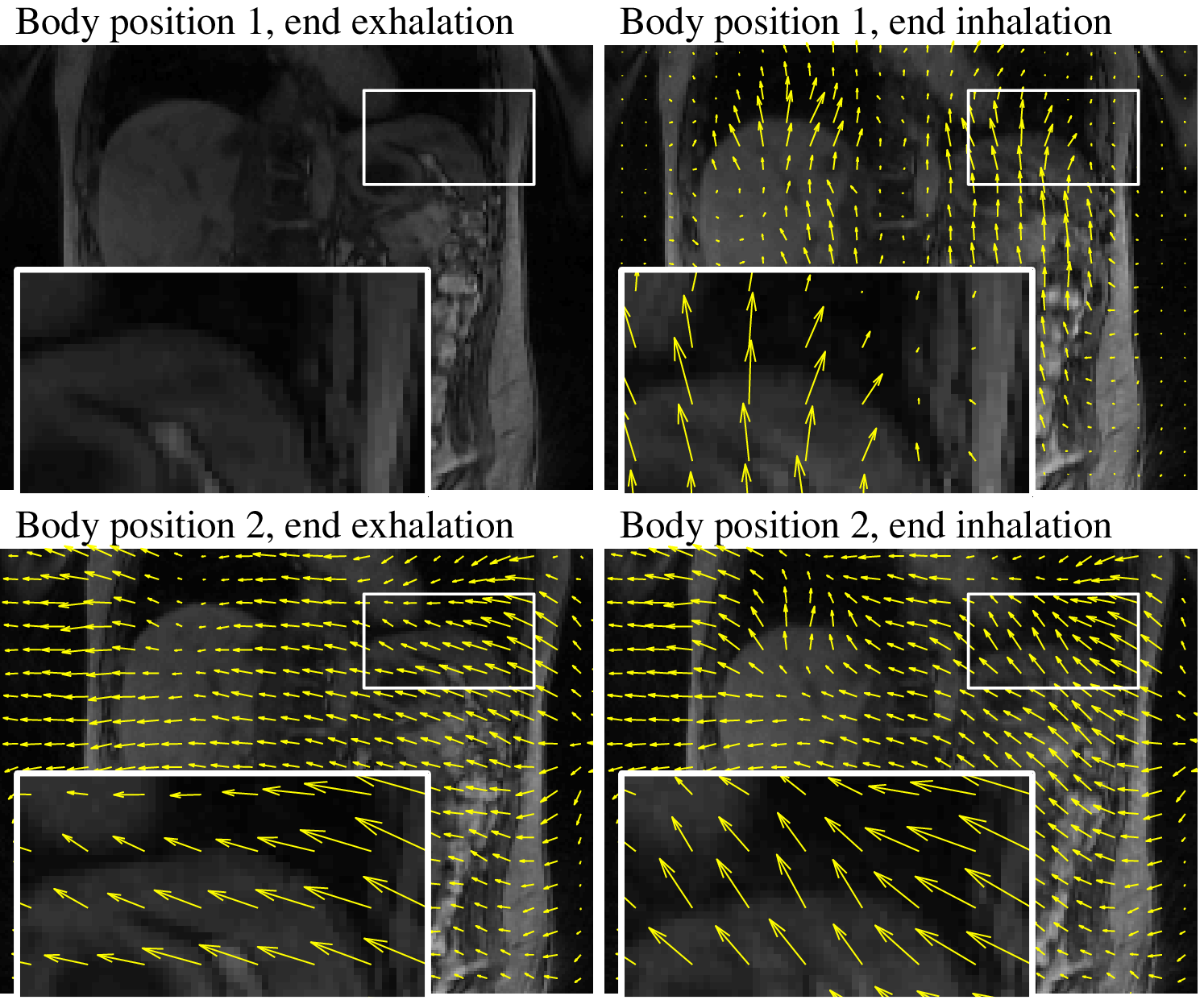}
\caption{\label{fig-pet-exp2-motion}Estimated motion field between bins from different bulk motion phases. The top left panel corresponds to the bin used as reference (first body position end inhalation). The top right panel shows the same body position at the end exhalation; the overlaid motion field exhibits mostly vertical displacement near the liver, corresponding to respiratory motion. The bottom row shows end inhalation and exhalation for the second body position (after bulk motion). Motion fields demonstrate the lateral displacement between body positions.}
\end{figure}

To account for the two body positions and for respiratory motion in PET
reconstruction, real-time MR images obtained by the proposed method were grouped
into 12 bins (6 bins for each body position) for motion field estimation.
Frames in the transition between the two bulk motion phases were excluded (a
total of 15 seconds were discarded). Motion was estimated between all bins and
the bin corresponding to the end-exhalation, which was used as reference bin.
Estimated motion fields are shown in figure~\ref{fig-pet-exp2-motion}. The top
left image shows motion caused by respiration, mostly visible as a vertical
displacement near the liver (shown with more details in the inset image). The
left column images show the bulk motion, which is mostly lateral. Finally, the
bottom right figure shows a combination of respiratory motion between end
inhalation and end exhalation and bulk motion.

\begin{figure}[htbp]
\centering
\includegraphics{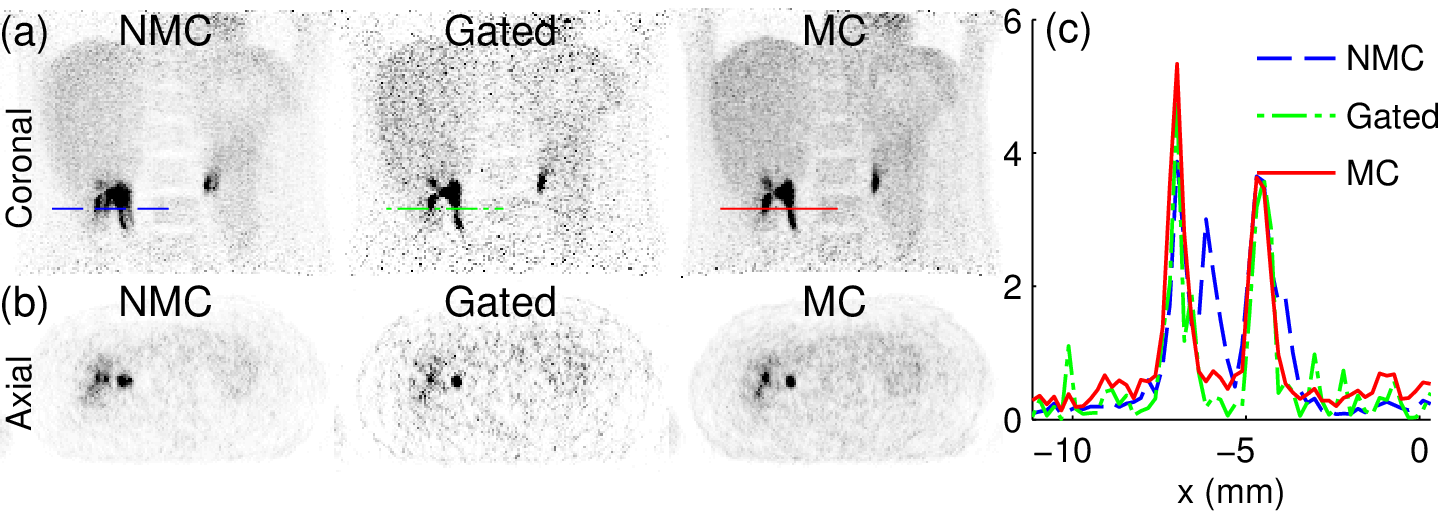}
\caption{\label{fig-pet-exp2-img}PET reconstructions for the bulk motion experiment using three different methods: reconstruction without motion correction (NMC), reconstruction from PET data corresponding to a single respiratory phase and body pose (Gated) and proposed motion-corrected reconstruction (MC). Profile plots through the right kidney are shown in (c).}
\end{figure}

Reconstructed PET images are shown in figure~\ref{fig-pet-exp2-img}.
Figures~\ref{fig-pet-exp2-img}(a) and~\ref{fig-pet-exp2-img}(b) show coronal
and axial slices using different reconstruction methods. Motion-blur is clearly
visible on the NMC reconstructions, primarily in the lateral direction,
corresponding to bulk motion but also in the vertical direction due to
respiratory motion. The gated reconstruction, which uses one sixth of the PET
counts at a single body position, shows sharper features but is severely
corrupted by noise. The proposed method compensates both respiratory and bulk
motion, significantly reducing motion-blur, while exhibiting a low noise level.
Figure~\ref{fig-pet-exp2-img}(c) shows line profiles through the kidney.
Without motion correction (NMC), the activity peak is lowered by motion blur.
Instead, two distinct peaks are visible, which correspond to the two bulk motion
phases. Gated and motion-corrected reconstructions both preserve the peak
activity, but gated reconstructions exhibit a high level of noise due to the
reduced amount of data used for reconstruction.

Evaluation measures are reported in table~\ref{tbl-pet-exp2-metrics}. The
table shows that motion corrected reconstruction leads to the highest CNR: a 83\%
improvement was observed over reconstruction without motion correction and 198\%
over gated reconstruction. The noise level in NMC and MC is similar (within
15\%) but the contrast is substantially improved by motion correction, while the
noise level in the gated reconstruction is three times higher leading to the low
CNR. For the TBR, the gated reconstruction achieves the highest ratio, because
gated reconstructions favor high contrast (at the expense of high noise). The
proposed motion compensation method approaches the gated TBR (25\% decrease) and
outperforms NMC (20\% increase).

\begin{table}[htbp]
\caption{\label{tbl-pet-exp2-metrics}Contrast-to-noise ratio (CNR) and target-to-background ratio (TBR) for kidney region of interest.  See figure~\ref{fig-pet-exp23-roi} for a view of the regions of interest.}
\centering
\begin{tabular}{lrrr}
 & NMC & Gated & MC\\
\hline
CNR & 42.07 & 25.88 & 77.26\\
TBR & 6.05 & 9.57 & 7.25\\
\end{tabular}
\end{table}

\subsection{Correction of irregular respiratory motion}
\label{results_experiment_3}
The second experiment was designed to evaluate the performance of the proposed
method in the case of irregular respiratory motion. The subject was instructed
to alternate between slow deep and fast shallow breaths throughout the 5 minutes
PET/MR acquisition.

\begin{figure}[htbp]
\centering
\includegraphics[width=0.9\textwidth]{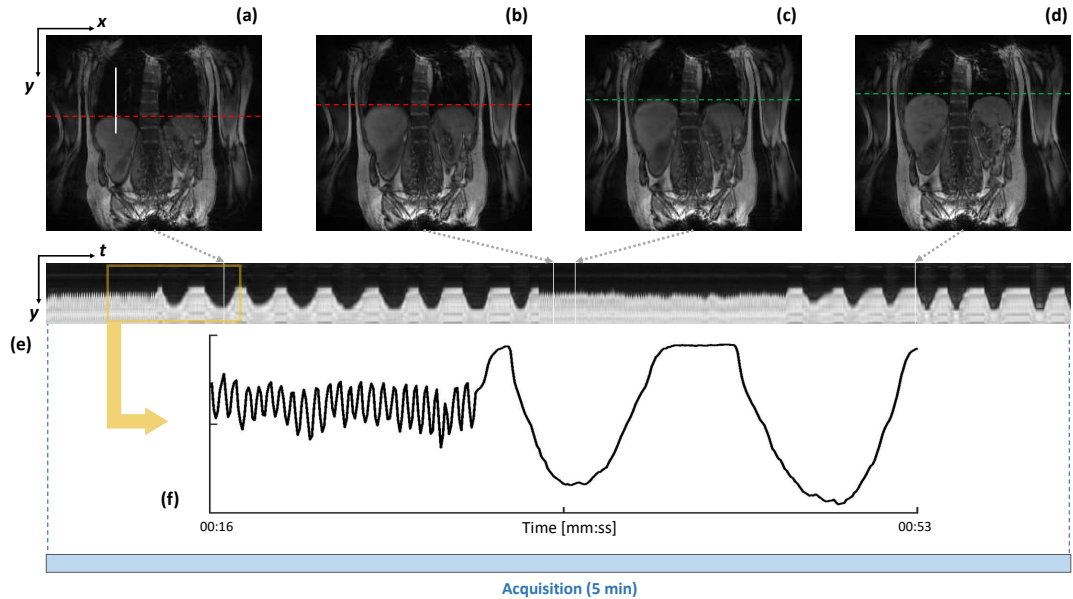}
\caption{\label{fig-mr-exp3-prof}Reconstructed MR images in the case of irregular respiratory motion. The real-time profile (e) is plotted for a few minutes of the experiment where the subject changes their breathing pattern from slow and deep to shallow and fast inspiration. Two representative images (a) \& (d) and (b) \& (c) are shown for each breathing pattern. Images (a) and (b) are shown for end-exhalation; (c) and (d) for end-inspiration. The red dashed-line indicates the top of the liver position for end-inhalation and the green dashed-line indicates the top of the liver position for end-exhalation for each breathing pattern.}
\end{figure}

MR images obtained by the proposed method are shown in
figure~\ref{fig-mr-exp3-prof}. The top row shows images at different frames:
two at the end of inhalation and two at the end of exhalation taken from
different breathing patterns (deep/shallow), respectively. The full extent of
the respiratory motion is captured and the images are artifacts-free.
Figures~\ref{fig-mr-exp3-prof}(e) and~\ref{fig-mr-exp3-prof}(f) show 1D
profiles of the image through the liver changing over time. Both the images and
the plot in figure~\ref{fig-mr-exp3-prof} clearly show the breathing patterns,
alternating between deep slow breaths and fast shallow ones. Based on the
reconstructed real-time MR images, 12 bins were determined through analysis of
the liver displacement in the MR images and were consequently used for motion
field estimation and motion corrected PET reconstruction. Sequences of MR
reconstructions are shown in Supplementary Material M2.

\begin{figure}[htbp]
\centering
\includegraphics{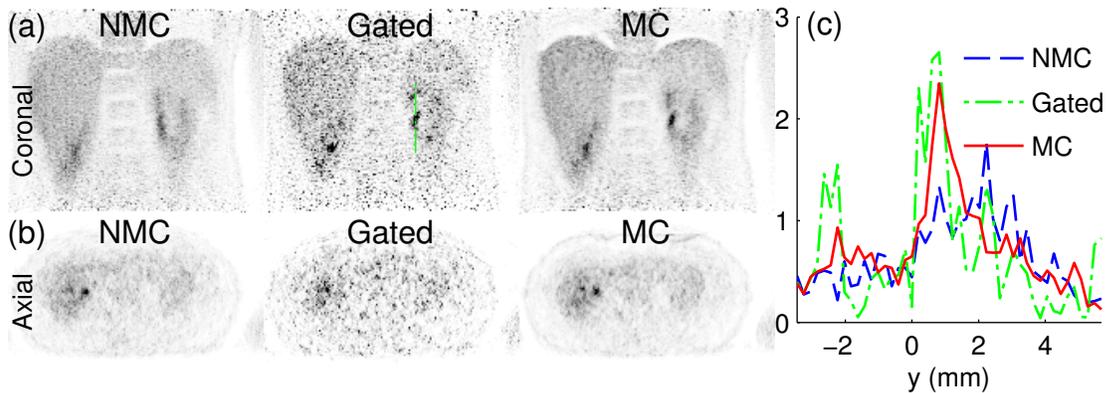}
\caption{\label{fig-pet-exp3-img}PET reconstructions for the irregular motion experiment using three different methods: reconstruction without motion correction (NMC), reconstruction from PET data corresponding to a single respiratory phase (Gated) and proposed motion-corrected reconstruction (MC). (c) shows profile plots through the right kidney.}
\end{figure}

Corresponding PET reconstructions are shown in figure~\ref{fig-pet-exp3-img}.
Images reconstructed without motion correction (NMC) exhibit blurring artifacts.
This is particularly visible on the left kidney (see the green line on the gated
coronal image) where the bright spot visible on other images is elongated in the
vertical direction, due to the large amplitude of the respiratory motion. The
gated reconstruction uses one sixth of the total number of counts and therefore
is degraded by noise, despite resulting in a sharper image. The proposed motion
correction method results in the best image quality, in terms of noise and
resolution. Corresponding line profiles are plotted in
figure~\ref{fig-pet-exp3-img}(c). The NMC peak is elongated along the y-axis,
due to the large extent of the mostly vertical respiratory motion. The Gated
line profile is sharper near its peak but has a large noise level. The proposed
MC method results in a good compromise between sharpness and low noise.
Contrast-to-noise and target-to-background ratios (defined in
Eq.~(\ref{eq_cnr}) and Eq.~(\ref{eq_tbr}) respectively) are reported in
table~\ref{tbl-pet-exp3-metrics} (regions of interest are shown in
figure~\ref{fig-pet-exp23-roi}(b)). Metrics show the superior performance of
the proposed motion correction method. The improvement in CNR is around 163\%
over NMC and over 200\% over gated reconstruction. The TBR for the proposed
method is within 15\% of the gated TBR and around 95\% larger than NMC.

\begin{table}[htbp]
\caption{\label{tbl-pet-exp3-metrics}Contrast-to-noise ratio and target-to-background ratio for kidney region of interest.  Regions are shown in figure~\ref{fig-pet-exp23-roi}.}
\centering
\begin{tabular}{lrrr}
 & NMC & Gated & MC\\
\hline
CNR & 17.52 & 15.28 & 46.18\\
TBR & 3.92 & 9.00 & 7.64\\
\end{tabular}
\end{table}

\section{Discussions}
\label{discussions}
We have demonstrated the performance of the proposed MR-based motion correction
for PET in two challenging cases: bulk motion and irregular respiratory motion.
The proposed subspace-based MR imaging method allows for reconstruction of
high-resolution 3D volumes at a rate of 9.5 volume/s, which enables accurate
motion field estimation even in the case of irregular motions. Another
important benefit of the proposed approach is the ability to perform informed
binning for PET motion correction, rather than relying on navigators or external
markers which offer limited information on the subject motion. With full
real-time volumetric MR images, detecting motion becomes straightforward, and
the process of determining an appropriate number of bins is greatly simplified.

The key assumption of the subspace-based imaging method is the low-rank property
of dynamic MR signals. We performed a simulation study to investigate this
property in the case of regular and irregular respiratory motion. Two phantoms
(shown in the Supplemental Material~M3 and~M4) were generated
using the XCAT software \citep{Segars2010} to simulate regular and irregular
respiratory motion. Respiratory and cardiac cycles were divided into
respectively 30 and 40 phases and 3D volumes were computed for each respiratory
and cardiac phase combination (i.e. 1200 volumes). A 4D (3D space + time)
phantom was then built by selecting and concatenating frame by frame the 3D
volumes based on simulated EKG and respiratory signals. Both the breathing
frequency and diaphragm expansion were varied while keeping a constant heart
rate in the simulation of the irregular respiratory motion. Each phantom
contained 6 respiratory cycles. The contrast was designed to simulate a
Balanced Steady-State Free Precession (bSSFP) signal for several compartments
such as fat, muscles, etc. using T1 and T2 values from the literature
\citep{Bojorquez2017}. SVD was then performed to investigate the effect of an
irregular respiratory pattern on the rank, and its corresponding approximation
error with low-rank truncation (see Supplemental Material figure~M5).
The decay of the calculated singular values from both phantoms was very similar,
indicating that the breathing pattern does not substantially affect the rank of
the data.

The proposed method utilizes an MR acquisition which fully overlaps with the PET
acquisition and provides real-time MR images for motion correction. The
proposed method can still have benefits for other commonly used acquisition
protocols. It is common in practice to reserve a first part of the PET
acquisition to perform MR motion field measurements and use the remaining PET
acquisition time to perform additional MR measurements (e.g. using T1 or T2
contrast sequences) that can be used for other diagnostic tasks
\citep{Petibon2019}. The proposed method can advantageously replace the motion
field measurement sequence, possibly reducing the acquisition time while
preserving image quality. A gating signal (e.g. navigator or external marker)
can then be used in subsequent MR sequences to select an appropriate bin for
each PET frame. Another approach is to integrate contrast sequences into the
motion field estimation sequence described in this paper. This is under
investigation and will be reported in separate publications.

The study reported in this paper has several limitations. First, the computation
time for the low-rank reconstruction with sparsely sampled non-Cartesian k-space
data could be a concern. The current MATLAB (The MathWorks, Inc., Natick,
Massachusetts, United States) implementation performs reconstruction of one
slice and one coil in around one hour. We anticipate that using a lower level
programming language and parallel computing devices (e.g. GPU) will help achieve
reasonable runtimes \citep{Wu2011b}.
Second, the proposed method does not have sufficient temporal resolution to
resolve the motion in the transition phase between the two bulk motion phases of
experiment~1 (figure~\ref{fig-mr-exp2-prof}(e)). The time-varying
profile plot shows that the image quality in the transition is severely
degraded. The corresponding list-mode data were excluded from the PET
reconstruction. Since the duration of the bulk motion was short, only about 5\%
of the list-mode data were discarded and thus should not be a significant
limitation. Third, this study focuses on demonstrating the feasibility of using
subspaced-based real-time MR for PET motion correction. We showed the
performance of our method in two cases (bulk motion and irregular respiratory
motion) from in vivo PET/MR experiments on a healthy subject. More subjects are
needed to fully evaluate the performance of the proposed method in clinical
settings.

\section{Conclusion}
\label{conclusion}
We proposed an MR-based method for PET motion correction using a subspace-based
real-time MR imaging for motion field estimation. We demonstrate the
feasibility of the proposed method using \textsuperscript{18}F-FDG-PET/MR
studies on a healthy subject. Our results show that the proposed method can
capture and correct for normal and irregular respiratory motions as well as
bulk body motion. The proposed method can be beneficial to a range of clinical
applications where irregular motion patterns are expected.

\section{Acknowledgments}
\label{acknowledgments}
This work was supported in part by the National Institutes of Health under award
numbers: T32EB013180, R01CA165221, R01HL118261, R21MH121812, R01HL137230 and
P41EB022544.

\bibliographystyle{jphysicsB}
\bibliography{bibliography}

\clearpage
\end{document}